\documentclass[twocolumn,showpacs,preprintnumbers,amsmath,amssymb]{revtex4}
\usepackage{graphicx}
\usepackage{dcolumn}
\usepackage{bm}
\begin{document}
%\preprint{APS/123-QED}
\title{Influence of temperature correlations on phase dynamics and kinetics
\\ of ultrathin lubricant film}
\author{A.~V.~Khomenko}
\email{khom@phe.sumdu.edu.ua}
\affiliation{Physical Electronics Department, Sumy State
University, 40007 Sumy, Ukraine}
\date{\today}
\begin{abstract}
The melting of ultrathin lubricant film is studied by friction between
atomically flat surfaces. The fluctuations of lubricant
temperature are taken into account defined by Ornstein-Uhlenbeck process.
The phase diagrams and portraits are calculated for the
cases of second-order and first-order transitions --- the melting of
amorphous and crystalline lubricants. It is shown that in
the first case the stick-slip friction domain appears,
dividing the regions of dry and sliding friction. In the second
case the three stick-slip friction domains arise characterized by
transitions between dry, metastable and stable sliding friction.
The increase of correlation time of the lubricant temperature
fluctuations leads to increasing of the frictional surfaces temperature
needed for realization of sliding friction.
The singular point, meeting the mode of dry friction, has
complex character of stability. The stationary states,
corresponding to the stable and metastable sliding friction,
are presented by the focus-type singular points.
\end{abstract}

\maketitle

\section{Introduction}\label{sec:level1}

The problems of the sliding friction continue to attract a considerable
attention due to vast application in engineering of flat solid surfaces
divided by thin film of lubricant \cite{Pers,Bowden,Yosh}.
It is found experimentally
that during the process of friction the liquid film becomes progressively
thinner, at first its physical properties change gradually (quantitatively),
and then the changes acquire the sharp (qualitative) character.
The qualitative changes consist of non-Newtonian mechanism of flow
and replacement of the ordinary melting by glass transition,
however, the lubricant film continues to behave itself like liquid.
In tribology such behaviour is called the "mixed lubrication", which
represents the intermediate mode of friction characterized by
transition from volume properties of lubricating material to
the boundary ones.

The boundary mode of friction, described in the proposed work, is
realized in the case of ultrathin lubricant films with thickness
less than four diameters of molecules at smooth surfaces or asperities,
high loads, and low shear rates. It is characterized by the following
changes of static (equilibrium) and dynamic properties of lubricant ---
simple unstructured Newtonian liquid \cite{Yosh}:
\begin{itemize}

\item non-fluidlike (non-Newtonian) properties: transition between liquid
and solid phases, appearance of new liquid-crystalline states,
epitaxially induced long-range ordering;

\item tribological properties: absence of flow until yield point or
critical shear stress reached, solidlike behaviour of lubricant
characterized by defect diffusion and dislocation motion, shear melting,
boundary lubrication.
\end{itemize}

Experiments with mica, silica, metal oxide, and surfactant monolayer surfaces,
between, which organic liquids and aqueous solutions were placed, have
shown that there are transformations between the different types
of dynamic phases during sliding \cite{Yosh}.
They manifest themselves in appearance of intermittent (stick-slip)
friction \cite{2lit,3lit,5lit,Aranson}, which is characterized by periodic
transitions between two or more dynamic states during the stationary
sliding and is the major reason for destruction and wear of frictional
elements. Thus, molecularly thin lubricant films undergo more than one
type of transition that results to existence of different types of
stick-slip motion.

In particular, the experimental investigation of rheological
properties of lubricant film were allowed to explain the nature of
above described anomalous features of boundary friction \cite{Yosh}.
Besides, with this aim the theoretical presentation of lubricating material
was used as the viscoelastic matter characterized by
heat conductivity. So, the observed phenomenology of ultrathin  lubricant
film is successfully described in \cite{Aranson} on the basis of
viscoelastic matter approximation and Ginzburg-Landau equation,
where order parameter defines the shear melting and freezing. The
phase diagram is calculated defining the domains of sliding, stick-slip,
and dry friction in the plane temperature --- film thickness.

Along the line \cite{Aranson},
within the framework of Lorenz model for description of viscoelastic matter
\cite{liq} the transition of
ultrathin lubricant film from solidlike into liquidlike state is shown to
occur due to both thermodynamic and shear melting. The cooperative
description of these processes is carried out as a result of
self-organization of the shear stress and stain, and the lubricant
temperature. The additive noises of above quantities are introduced for
building the phase diagrams, where fluctuations intensities and frictional
surfaces temperature define the domains of sliding, stick-slip, and dry
friction \cite{pla}. In reference~\cite{physa_soc} the conditions are found at
which the stick-slip friction regime corresponds to the intermittency mode
inherent in self-organized criticality phenomenon.

However, in spite of the fact that, as a rule, temperature determines the
lubricant state, the question remains opened on influence of its correlated
fluctuations on the process of friction. Here, it is shown that internal
fluctuations of temperature, which are described by the Ornstein-Uhlenbeck
process, result in complication of dynamic phase diagram. The shear stress
distribution function, the phase diagrams and portraits are calculated for the
cases of second-order and first-order transitions --- the melting of amorphous
and crystalline lubricant (Secs.~\ref{sec:level2}~and~\ref{sec:level3}),
respectively. The different stick-slip friction modes are predicted to occur.
The influence of correlation time value of the studied fluctuations is
described on the phase diagram. The investigation of kinetic modes of boundary
friction is carried out using the phase plane method. The self-similar phase
dynamics and kinetics of lubricant film are investigated.

\section{Continuous transition}\label{sec:level2}

In the previous work \cite{liq} on the basis of rheological description of
the viscoelastic medium the system of kinetic
equations has been obtained, which define the mutually coordinated evolution
of the shear components of the stress $\sigma$ and the strain
$\varepsilon$, and the temperature $T$ in ultrathin lubricant film during
friction between atomically flat mica surfaces. Let us write these
equations using the measure units
\begin{eqnarray}
\sigma_{s}&{=}&\left( {\rho c_{\upsilon}\eta_{0}T_{c} \over \tau
_{T}}\right)^{1/2}, \nonumber \\ \varepsilon_{s}&{=}&
{\frac{\sigma_{s}}{G_{0}}}{\equiv} \left( {\frac{\tau _{\varepsilon }}{\tau
_{T}}}\right) ^{1/2}\left( {\frac{\rho c_{\upsilon}T_{c}\tau _{\varepsilon }
}{\eta_{0}}}\right)^{1/2},\quad T_{c} \label{1aa} \end{eqnarray} for variables
$\sigma$, $\varepsilon $, $T$, respectively, where $\rho$ is the mass density,
$c_v$ is the specific heat capacity, $T_{c}$ is the critical temperature,
$\eta_{0} \equiv \eta(T=2T_{c})$ is the typical value of shear viscosity
$\eta$, $\tau_{T}\equiv\rho l^2 c_{\upsilon}/\kappa$ is the time of heat
conductivity, $l$ is the scale of heat conductivity, $\kappa$ is the heat
conductivity constant, $\tau_{\varepsilon}$ is the relaxation time of matter
strain, $G_{0}\equiv \eta _{0}/\tau _{\varepsilon}$:
\begin{eqnarray}
&&\tau _{\sigma}\dot{\sigma}=-\sigma + g\varepsilon , \label{eq2} \\
&&\tau_{\varepsilon }\dot{\varepsilon}=-\varepsilon + (T-1)\sigma ,
\label{eq3} \\ &&\tau _{T}\dot{T}=(T_{e}-T) - \sigma \varepsilon +
\sigma ^{2}  + \lambda(t) . \label{eq4} \end{eqnarray}
Here the stress relaxation time $\tau_{\sigma}$, the temperature $T_{e}$
of atomically flat mica friction surfaces, and the
constant $g=G/G_{0}$ are introduced, where $G$ is the lubricant shear
modulus. Replacement of $\varepsilon / \tau_{\sigma}$ by
$\partial \varepsilon / \partial t$
reduces Eq.~(\ref{eq2}) to the Maxwell-type equation for
viscoelastic matter approximation that
is widely used in the theory of boundary friction \cite{Pers}.
The relaxation behaviour of viscoelastic lubricant during the process of
friction is described also by Kelvin-Voigt equation~(\ref{eq3})
\cite{liq,voigt}. It takes into account the dependence of the shear
viscosity on the dimensionless temperature $\eta = \eta_{0}/(T-1)$
\cite{marvan}. Equations~(\ref{eq2}) and (\ref{eq3}) represent jointly
the new rheological model. It is worth noting that rheological properties
of lubricant film are investigated experimentally for
construction of phase diagram \cite{Yosh}.
Equation~(\ref{eq4}) represents the heat conductivity expression,
which describes the heat transfer from the friction surfaces to the
layer of lubricant, the effect of the dissipative heating
of a viscous liquid flowing under the action of the stress, and the reversible
mechanic-and-caloric effect in linear approximation.
Equations ~(\ref{eq2}) -- (\ref{eq4})
coincide with the synergetic Lorenz system formally
\cite{Haken,zhetph}, where the shear stress
acts as the order parameter, the conjugate field
is reduced to the shear strain, and the temperature is the control
parameter. As is known this system can be used for description of
the thermodynamic phase and the kinetic transitions.

The purpose of this work is to study the phase dynamics and
kinetics of boundary friction at introduction into
Eq.~(\ref{eq4}) the stochastic source $\lambda(t)$ representing the
Ornstein-Uhlenbeck process:
\begin{eqnarray}
\langle\lambda(t)\rangle = 0,\quad
\langle\lambda(t)\lambda(t')\rangle=
\frac{I}{\tau_\lambda}\exp{\left(-\frac{|t-t'|}{\tau_\lambda}\right)},
\label{Teq7}\end{eqnarray}
where $I$ is the temperature fluctuations intensity,
$\tau_{\lambda}$ is the time of their correlation.

Following \cite{stat}, let us define the physical sense of value $I$.
The time correlation of dimensionless temperature of lubricant is
determined by the average value of product
\begin{equation}
\label{1a}
\varphi(\tau)= \langle \Delta T(t)\Delta T(t+\tau)\rangle,
\end{equation}
where $\Delta T(t)$ is the difference between the current and the
average temperature. Within the framework of assumption on the
quasistationarity of fluctuations $\Delta T$ the function $\varphi(\tau)$
accepts the form:
\begin{equation}
\label{2a}
\varphi(\tau)= \langle (\Delta T)^2 \rangle \exp(-\zeta |\tau|).
\end{equation}
Here $1/\zeta$ fixes the relaxation time for setting of equilibrium.
Supposing that in formulas (\ref{1a}) and (\ref{2a}) the moments of
time are connected by equality $t+\tau = t'$, we get:
\begin{equation}
\label{3a}
\varphi(t'-t)= \langle \Delta T(t)\Delta T(t')\rangle =
\langle (\Delta T)^2 \rangle \exp(-\zeta |t'-t|).
\end{equation}
Then, substitution of average square of temperature fluctuations
$\langle (\Delta T)^2 \rangle = T^2/c_v$ in expression (\ref{3a}) gives
\begin{equation}
\label{5a}
\langle \Delta T(t)\Delta T(t')\rangle = \frac{T^2}{c_v}
\exp(-\zeta |t'-t|).
\end{equation}
Comparing this formula with the second equality (\ref{Teq7}), we obtain:
\begin{equation}
\label{result}
\lambda(t)= \Delta T(t), \quad \tau_\lambda = \frac{1}{\zeta}, \quad
I = \frac{T^2}{c_v\zeta}.
\end{equation}
Consequently, the value of noise intensity $I$ is defined by the temperature
and the heat capacity $c_v$ of lubricant. At the first glance it can appear,
that in the phase diagrams $T_e(I)$ presented further, the every value
$I$ meets the unique $T_e$, and instead of regions, there
the curve has to exist in pointed out coordinates. The motion
along this curve describes the evolution of the system.
However, it does not so, since parameter $T_e$ represents the
thermostat temperature whose arbitrary change does not determine
the temperature $T$ uniquely. It is necessary only to suppose that
as a result of system's self-organization the defined value $T_e$
can meet the manifold of values $T$ and, in accordance with
(\ref{result}), the intensities $I$ in the different moments of time.
Besides, the variation of $c_v$ leads to the change of $I$ in the
course of time. Thus, it implies the existence of
phase diagram. It is possible also to change the intensity $I$
due to the arbitrary choice of parameter $\zeta$ characterizing
the concrete system. The last means that the systems exist, where
the noise does not influence substantially on their time behaviour,
and the ones, where fluctuations influence critically.

However, here we do not restrict ourselves by such approach and
understand temperature fluctuations in more wide sense.
It is related to that, as a rule, the thermal influence is caused
by different external stochastic sources.
Besides, it is possible to interpret
the noise with the help of fluctuations $T$, as well as of $T_e$.
This presentation of noise is often used for its modeling in real
systems \cite{gard}.

In reference~\cite{liq} a melting of ultrathin lubricant film by friction
between atomically flat mica surfaces has been represented as a result of
action of spontaneously appearing shear stress leading to the  plastic
flow. This is caused by the heating of friction surfaces above
the critical value $T_{c0}=1+g^{-1}$. Thus, according to such approach
the studied solidlike-liquidlike transition of lubricant film
occurs due to both thermodynamic and shear melting.
The initial reason for this self-organization process
is the positive feedback of $T$ and $\sigma$ on
$\varepsilon$ [see Eq.~(\ref{eq3})] conditioned by the temperature
dependence of the shear viscosity leading to its divergence.
On the other hand, the negative feedback of $\sigma$ and $\varepsilon$
on $T$ in Eq.~(\ref{eq4}) plays an important role since it ensures the system
stability.

According to this approach the lubricant represents a strongly viscous liquid
that can behave itself similar to the solid --- has a high effective
viscosity and still exhibits a yield stress \cite{Yosh,voigt}.
Its solidlike state corresponds to the shear stress $\sigma=0$
because Eq.~(\ref{eq2}), describing the elastic
properties at steady state $\dot\sigma=0$, falls out of consideration.
Equation (\ref{eq3}), containing the viscous stress, reduces to
the Debye law describing the rapid relaxation of the shear strain
during the microscopic time $\tau_{\varepsilon} \approx a/c \sim 10^{-12}$ s,
where $a\sim 1$~nm is the lattice constant or the intermolecular distance
and $c\sim 10^3$~m/s is the sound velocity. At the same time
the heat conductivity equation (\ref{eq4}) takes on the form of
simplest expression for temperature relaxation that does not contain the
terms representing the dissipative heating and the mechanic-and-caloric
effect of a viscous liquid.

Equation~(\ref{eq3}) describes the flow of lubricant with velocity
$V=l\partial\varepsilon/\partial t$ due to action of appearing
viscous shear stress.
Moreover, in accordance with Ref.~\cite{Aranson} in the absence of shear
deformations the temperature mean-square
displacement is defined by equality $\langle u^2\rangle=T/Ga$.
The average shear displacement is found from the relationship
$\langle u^2\rangle =\sigma^2a^2/G^2$. The total mean-square displacement
represents the sum of these expressions provided that the thermal
fluctuations and the stress are independent.
Above implies that the transition of lubricant from solidlike to
fluidlike state is induced both by heating and under influence of stress
generated by solid surfaces at friction. This agrees
with examination of solid state instability within the framework
of shear and dynamic disorder-driven melting representation in absence
of thermal fluctuations.
It is assumed that the film becomes more liquidlike and the friction
force decreases with the temperature growth due to decreasing activation
energy barrier to molecular hops.
Besides, the friction force decreases with
increasing velocity at the contact $V=l\partial\varepsilon/\partial t$
because the latter leads to the growth of the shear stress $\sigma$
according to the Maxwell stress - strain $\varepsilon$ relationship:
$\partial\sigma /\partial t {= -}\sigma / \tau_\sigma {+} G\partial\varepsilon/
\partial t.$

This work is devoted to study of the stochastic source $\lambda(t)$
influence on the evolution of the stress $\sigma(t)$.
In accordance with experimental data for the organic lubricant
\cite{Yosh,pla} the stress relaxation time at normal pressure is
equal to $\tau_{\sigma} \sim 10^{-10}$~s. Since ultrathin lubricant film
consists of less than four molecular layers, the temperature relaxes
to the value $T_e$ during the time satisfying the
inequality $\tau_T \ll \tau_{\sigma}$. Therefore, we will suppose
that conditions are fulfilled
\begin{equation}
\label{Teq8}
\tau_{\sigma} \approx \tau_{\varepsilon} \gg \tau_{T},
\end{equation}
at which lubricant temperature $T$ follows the change of the shear
components of stress $\sigma $ and strain $\varepsilon$. Then,
it is possible to select a small parameter and to put
$\tau_T \dot T \simeq 0$ in Eq.~(\ref{eq4}). As a result, we obtain the
expression for temperature:
\begin{equation}\label{Teq9}
T = T_e  - \sigma\varepsilon +\sigma^2 + \lambda(t).
\end{equation}

Let us give for the system (\ref{eq2}), (\ref{eq3}), and (\ref{Teq9}) the
more simple form, reducing it to the single equation for the shear stress
$\sigma(t)$.
For this purpose it is necessary to express $\varepsilon$ and $T$
via $\sigma$. Differentiating with respect to time the equation
for strain $\varepsilon$ that is obtained from (\ref{eq2}),
we get equation for $\dot{\varepsilon}$.
Substitution of these expressions for $\varepsilon$, $\dot{\varepsilon}$ and
the equality (\ref{Teq9}) in (\ref{eq3}) gives evolution equation in
canonical form of equation for nonlinear stochastic oscillator
of the van der Pole generator type:
\begin{equation}\label{Teq12}
m\ddot\sigma+\gamma(\sigma)\dot\sigma = f(\sigma)+ \phi(\sigma)\lambda(t),
\end{equation}
where the coefficient of friction $\gamma$, the force $f$, the amplitude
of noise $\phi$, and the parameter $m$ are defined by expressions
\begin{eqnarray}
\gamma(\sigma) &\equiv& \frac{1}{g}
\left[\tau_\varepsilon+\tau_\sigma(1+\sigma^2) \right],\nonumber \\
f(\sigma)&\equiv& \sigma\left(T_e-1-g^{-1}\right) -
\sigma^3\left(g^{-1}-1\right), \nonumber \\ \phi (\sigma) &\equiv& \sigma,
\quad m \equiv \frac{\tau_\sigma \tau_\varepsilon}{g}. \label{Teq13}
\end{eqnarray}
Let us find the distribution function of the stress $\sigma$.
With this aim, we will use the method of effective potential
\cite{Shapiro}, \cite{Kharch}. As a result, the
Fokker-Planck equation is obtained
\begin{equation}\label{dFok_Plank}
\frac{\partial P}{\partial t}=-\frac{\partial }{\partial
\sigma }\left(D^{(1)}P\right)+
\frac{\partial}{\partial\sigma}\left(D^{(2)}
\frac{\partial P}{\partial\sigma}\right).
\end{equation}
It is expressed in terms of coefficients
\begin{eqnarray}
D^{(1)}&=&\frac{1}{\gamma}\left\{f {-} I\phi^2\frac{\partial
\gamma^{-1}}{\partial\sigma}{-}\phi \frac{\partial \phi}{\partial\sigma}
\left(\frac{2 I}{\gamma}{+} I \tau_\lambda\right) \right\},
\label{dD1} \\
D^{(2)}&=&\frac{\phi^2}{\gamma}\left(\frac{I}{\gamma}+2 I\tau_\lambda\right).
\label{dD2}
\end{eqnarray}
In the stationary case the solution of Eq.~(\ref{dFok_Plank}) leads
to the distribution
\begin{equation}\label{dpdf}
P(\sigma)=\mathcal{Z}^{-1} \exp\left\{ -E(\sigma) \right\},
\end{equation}
which is fixed by effective potential
\begin{equation}\label{potential}
E(\sigma) = -\int\limits_{0}^\sigma \frac{D^{(1)}(x)}{D^{(2)}(x)}{\rm d}x
\end{equation}
and normalization constant
\begin{equation}
\mathcal{Z}=\int\limits_{0}^{\infty}{\rm d}\sigma
\exp\left(\int\limits_{0}^\sigma\frac{D^{(1)}(x)}{D^{(2)}(x)}
{\rm d}x\right).
\end{equation}
The stationary shear stress is found from the extremum condition of
distribution (\ref{dpdf})
\begin{equation}
\frac{D^{(1)}(\sigma)}{D^{(2)}(\sigma)}=0. \label{extr}
\end{equation}
According to Fig.~1 the distribution (\ref{dpdf}) has pronounced
maximums whose positions are determined by the set of parameters
$\tau_\sigma$, $\tau_\varepsilon$, $\tau_\lambda$, $g$, $I$, and $T_e$.
At the small values of friction surfaces temperature $T_e$
a single maximum is realized at the point $\sigma=0$ meeting the dry friction
mode (curve 1). With $T_e$ growth the two maximums appear
at points $\sigma = 0$ and $\sigma \ne 0$  (curve 2),
the first of them corresponds to the dry friction, the
second one --- to the sliding. Here the stick-slip friction mode,
characterized by transitions between the indicated stationary
regimes, is realized. With the further growth of $T_e$ the
zero maximum of $P(\sigma)$ disappears and
the maximum at $\sigma \ne 0$ remains only (curve 3),
i.e., lubricant becomes liquidlike.

Supposing in (\ref{extr}) $\sigma=0$, we find the critical value
of friction surfaces temperature
\begin{equation} \label{Teq35}
T_{e0} = \frac{1+g}{g}
+\left(\tau_\lambda +\frac{2g}{\tau_\varepsilon+\tau_\sigma}\right)I
\end{equation}
providing transition to the sliding mode.
It is seen that $T_{e0}$ grows at the increase of noise
intensity $I$ and correlation time $\tau_\lambda$.
The values of relaxation times of the shear stress and strain influence
by reverse manner. The domains of dry (DF), sliding (SF), and
stick-slip (SS) modes of friction are realized in the phase
diagram presented in Fig.~2. The growth of correlation
time $\tau_{\lambda}$ results in the increase of value $T_e$,
corresponding to the tricritical point $T$ at defined intensity $I$.
Thus, the region of dry friction broadens, while
sliding and stick-slip friction becomes more hardly realized.

Apparently, the increase of the sheared
surfaces temperature $T_e$ transforms lubricant
to the sliding friction mode. It can be understood considering
Eq.~(\ref{Teq12}) that describes the damping oscillations.
Here the surfaces temperature is included only in expression
for driving force $f$, which increases with growth of $T_e$.
As is known, the liquid state can correspond
to the oscillation mode with large amplitude, but the solid state
can not. With the increasing value of
effective force in (\ref{Teq13}) the amplitude of oscillations increases,
and longer oscillation process is realized to the moment of setting
of the certain friction mode in the system.

For studying the dynamics of change of friction modes
it is enough to represent the distribution $P(\sigma)$
by position of its maximum $\tilde\sigma$.
This is achieved by the use of the path integrals formalism
\cite{Zinn_Justin}, within the framework of which the extreme values
$\tilde\sigma=\tilde\sigma(t)$ of the
initial distribution function (\ref{dpdf})
evolve in accordance with the effective distribution
\begin{equation}\label{Teq28}
\Pi \{ \dot{\tilde\sigma},\tilde\sigma \} \propto\exp\left(-
\int\Lambda(\dot{\tilde\sigma},\tilde\sigma,t){\rm d}t \right).
\end{equation}
Here the Onsager-Machlup function $\Lambda$, acting as the Lagrangian
of Euclidean field theory, is the subject for determination.

Equation (\ref{dFok_Plank}) can be transformed to the form:
\begin{eqnarray} \label{dFok_Plank_ito}
\frac{\partial P}{\partial t}=-\frac{\partial}{\partial \sigma
}\left[\left(D^{(1)} + \frac{{\rm d} D^{(2)}}{{\rm d}\sigma}\right) P\right] +
\frac{\partial^2}{\partial\sigma^2} \left(D^{(2)} P\right).
\end{eqnarray}
For finding of the $\Lambda(\dot{\tilde\sigma},\tilde\sigma,t)$ dependence
we write down the differential Langevin equation
\begin{equation}\label{Teq25}
{\rm d}\tilde\sigma = \left(D^{(1)}+
\frac{{\rm d}D^{(2)}}{{\rm d}\sigma}\right){\rm d}t +
\sqrt{2D^{(2)}}{\rm d}W(t)
\end{equation}
corresponding to the Fokker--Planck equation (\ref{dFok_Plank_ito})
\cite{Zinn_Justin}. Here the stochastic differential ${\rm d}W(t)$
represents the Winner process with the properties of white noise:
\begin{equation} \label{Teq24}
\langle{\rm d}W(t)\rangle=0,\quad\langle({\rm d}W(t))^2\rangle={\rm d}t.
\end{equation}
The feature of stochastic equations is that
differential ${\rm d}W(t)$ can not be obtained by the simple division
of Eq.~(\ref{Teq25}) by $\sqrt{2D^{(2)}}$. With this aim,
it is necessary to pass from the random process $\tilde\sigma(t)$ to the
white noise $x(t)$ related with the initial Jacobian
${\rm d}x/{\rm d}\tilde\sigma = \left(2 D^{(2)}\right)^{-1/2}$. Then,
substitution of Eq.~(\ref{Teq25}) into the Ito stochastic differential
\begin{equation} \label{Teq26}
{\rm d}x = \frac{{\rm d}x}{{\rm d}\tilde\sigma}{\rm d}\tilde\sigma +
\frac{1}{2}\frac{{\rm d}^2x}{{\rm d}\tilde\sigma^{2}}
({\rm d}\tilde\sigma)^2,
\end{equation}
taking into account (\ref{Teq24}), leads to the expression
\begin{eqnarray} \label{Teq26a}
{\rm d}x &=& \left(\frac{{\rm d}x}{{\rm d}\tilde\sigma}\left( D^{(1)} +
\frac{{\rm d}D^{(2)}}{{\rm d}\tilde\sigma}\right)+ \frac{{\rm d}^2x}{{\rm
d}\tilde\sigma^{2}} D^{(2)}\right){\rm d}t \nonumber
\\&+&\frac{{\rm d}x}{{\rm d}\tilde\sigma}\sqrt{2 D^{(2)}}{\rm d}W(t).
\end{eqnarray}
Here the terms are neglected whose order exceeds
$\left({\rm d}W(t)\right)^2$. After reverse transition from the white noise
$x(t)$ to the initial process $\tilde\sigma(t)$ the equality
\begin{equation}\label{Teq27}
\frac{{\rm d}W(t)}{{\rm d}t}= \frac{\dot{\tilde\sigma}}{\sqrt{2D^{(2)}}}-
\frac{2D^{(1)} + (D^{(2)})^\prime}{2\sqrt{2D^{(2)}}}
\end{equation}
is obtained, where the stroke stands for differentiation with respect
to $\tilde\sigma$. Plugging this expression into Gaussian
\begin{eqnarray}
\Pi\propto\exp\left\{-
\frac{1}{2}\int\left(\frac{{\rm d}W(t)}{{\rm d}t}\right)^2{\rm d}t\right\}
\nonumber
\end{eqnarray}
and comparing with (\ref{Teq28}), we arrive at Lagrangian
\begin{equation}
\label{Teq29}
\Lambda = \frac{1}{4}\frac{{\dot{\tilde\sigma}^2}}{D^{(2)}} - U
\end{equation}
with potential energy
\begin{equation}\label{Teq30_old}
U = -\frac{(2D^{(1)}+(D^{(2)})^\prime)^2}{16D^{(2)}}.
\end{equation}
It is substantial that such form of potential energy $U$
does not coincide with the effective potential
(\ref{potential}). Consequently, for further
consideration it is necessary to replace the expression $U$
(\ref{Teq30_old}) by $E(\sigma)$ (\ref{potential}).
In this case, Eq.~(\ref{Teq29}) describes the system behaviour
in accordance with distribution (\ref{dpdf}) obtained above.

The system kinetics is defined by Euler--Lagrange equation
\begin{equation}\label{Teq36}
\frac{\partial\Lambda}{\partial\tilde\sigma} -
\frac{\partial}{\partial t}\frac{\partial\Lambda}{\partial\dot{\tilde\sigma}}
=\frac{\partial R}{\partial\dot{\tilde\sigma}}.
\end{equation}
Within the white noise presentation the
dissipative function has the simplest form
$R ={\dot x}^2/2$ and it is transformed to
\begin{equation}\label{Teq36aa}
R=\frac{\dot{\tilde\sigma}^2}{4 D^{(2)}}
\end{equation}
with transition to the variable $\dot{\tilde\sigma}=(2D^{(2)})^{1/2}\dot x$.
Substituting in (\ref{Teq36}) the equalities (\ref{Teq29}), (\ref{potential}),
(\ref{Teq36aa}), we arrive at differential equation
\begin{equation}\label{Teq37}
\ddot{\tilde\sigma} + \dot{\tilde\sigma}^2\frac{(D^{(2)})^\prime}{2D^{(2)}}+
\dot{\tilde\sigma} - 2\frac{D^{(1)}}{D^{(2)}}D^{(2)} = 0.
\end{equation}
At the steady state supposing in (\ref{Teq37}) $\dot{\tilde\sigma}= 0$,
the extremum condition (\ref{extr}) of distribution (\ref{dpdf}) is
obtained. Note that the distribution maximum corresponds
to the minimum of effective potential,
and the distribution minimum --- to its maximum.

Consider kinetics of the system using the phase plane method for analysis
of Eq.~(\ref{Teq37}). The phase portraits are presented in Fig.~3
corresponding to the curves of Fig.~1.

The region of dry friction (DF) (Fig.~3a) is characterized by the
presence of the one singular point $D$,
which corresponds to the maximum of probability
$P(\sigma)$ at $\sigma = 0$. This point is non-standard and it requires
interpretations. It is located at the origin of coordinates, and phase
trajectories are curved around it, so that the system never comes to the value
$\sigma = 0$, i.e., this point is not stationary. Consider the system behaviour
at the arbitrary initial condition. According to the phase trajectories the
system evolves to the zero value of stress. Thus, if at initial conditions the
growth rate of stress is positive it decreases to the zero at first (during
this time the stress increases). Then, the stress value decreases
asymptotically to the zero with the increase of its decrease rate. Presumably,
the described situation meets the mode of explosive amorphization, at which the
system transforms very rapidly in amorphous solidlike state. The circumstance
that zero stress is never achieved testifies to divergence of the
probability $P(\sigma)$ at point $\sigma = 0$.
This is related with infinite growth of decrease
rate of stress variation at going of the system to $\sigma = 0$. Let us assume
that the system has reached
the point, at which decreasing rate becomes critical. In
such case the increase of stress value is expected, which is accompanied by the
change of rate sign and transition of the system in the positive phase plane
domain. Further, again the decreasing of stress occurs, and the described
situation repeats oneself. Alternatively, the sign of stress variation rate
does not change, and only its decreasing takes place. This moves the system on
a neighboring phase trajectory, along which it approaches to the zero stress
more rapidly. As a result, in the course of time the oscillation mode of dry
friction is set in the vicinity of point $\sigma = 0$ at the arbitrary initial
conditions. Thus, the oscillations amplitude is small, and lubricant has
solidlike structure.

The phase portrait of the system, characterizing the region of stick-slip
friction (SS), is shown in Fig.~3b. The three singular points appear here:
$D$, the saddle $N$, and the stable focus $F$.
As well as above, the point $D$
is realized at the origin of coordinates and it corresponds to the dry
friction mode. The saddle $N$ meets the
minimum of $P(\sigma)$ and it is unstable
stationary point. It is worth noting that at the initial value of the
shear stress on the right-hand side from point $N$ and $\dot\sigma=0$,
the sliding mode of friction is set in the system during time. If
the initial value of stress appears on the left-hand side from saddle $N$,
the dry friction is set in similar case. Thus, the point $N$
separates two maximums of distribution function $P(\sigma)$.
The focus $F$ corresponds to the non-zero maximum
of stress distribution function,
i.e., it describes the liquidlike state of lubricant.
The damping oscillations, corresponding to this point, mean
that lubricant becomes more liquidlike and more solidlike periodically.
However, the stable sliding friction is set always. Presumably,
these oscillations are conditioned by the presence of noise.

The phase portrait, meeting the sliding friction (SF) (Fig.~3c)
characterized by single non-zero maximum of distribution
function $P(\sigma)$, has one singular point
--- the stable focus $F$. The stability of sliding friction
is confirmed also by large overexpansion
of phase trajectories near $F$ along axes of ordinates and abscissas.
However, it is apparent that at the initial large value
$\sigma$ the system does not reach the point $F$,
and approaches asymptotically to the zero stress value.
This circumstance implies that conditions can be realized,
at which the system will be near the regime of dry friction.
As described above, at reaching the value of critical variation rate
of stress its sign changes and becomes positive. It is seen from the phase
portrait that in such case the system also passes to the mode of
stable sliding friction.

\section{Influence of deformational defect of modulus}\label{sec:level3}

Actually, the shear modulus introduced (in terms of the relaxation time
$\tau_\sigma$) in Eq.~(\ref{eq2}), depends on the stress value.
This leads to the transition of the elastic deformation mode to
the plastic one. It takes place at
characteristic value of the stress $\sigma_p$, which
does not exceed the value $\sigma_s$ (in other case the
plastic mode is not manifested).
For consideration of deformational defect of the modulus
we will use $\tau_\sigma(\sigma)$ dependence proposed in \cite{liq},
instead of $\tau_\sigma$. As a result, Eq.~(\ref{eq2}) takes the form:
\begin{equation}
\tau_{p}\dot\sigma = - \sigma\left(1+\frac{\theta^{-1}-1}{1+\sigma/\alpha}
\right) + g_\Theta\varepsilon, \label{eq2n}
\end{equation}
where the relaxation time for the plastic mode
$\tau_p = \eta_\sigma / \Theta$ is introduced
($\eta_\sigma \equiv \tau_\sigma G$ is the effective viscosity,
$\Theta$ is the hardening factor), $\theta = \Theta / G <1$
is the parameter describing the ratio of tilts of
the plastic and the Hookean sections of the deformation curve,
$g_\Theta = G^2/\Theta G_0$ and $\alpha=\sigma_p/\sigma_s$ are the constants.
Then, within the framework of approximation (\ref{Teq8}) the system
(\ref{eq2n}), (\ref{eq3}), and (\ref{eq4}), as well as above, is reduced
to equation (cf. (\ref{Teq12})):
\begin{equation}\label{Teq12n}
m\ddot\sigma+\gamma(\sigma)\dot\sigma = f(\sigma)+ \phi(\sigma)\lambda(t),
\end{equation}
where the coefficient of friction $\gamma$, the force $f$, the
amplitude of noise $\phi$, and the parameter $m$ are defined by expressions
\begin{eqnarray}
&&\gamma(\sigma)\equiv \frac{1}{g}\left[\tau_\varepsilon
\left(1+\frac{\theta^{-1}-1}{(1+\sigma/\alpha)^2}\right)+\tau_p(1+\sigma^2)
\right], \nonumber \\ &&f(\sigma)\equiv \sigma\left[T_e-1-\frac{1}{g}
\left(\frac{\theta^{-1}+\sigma/\alpha}{1+\sigma/\alpha}\right)\right] \nonumber
\\ &&- \sigma^3\left[\frac{1}{g}\left(\frac{\theta^{-1}+\sigma/\alpha}
{1{+}\sigma/\alpha}\right){-}1\right], \phi(\sigma){\equiv}\sigma, m {\equiv}
\frac{\tau_p \tau_\varepsilon}{g_\Theta}. \label{ggn}
\end{eqnarray}
According to the effective potential method \cite{Shapiro},
\cite{Kharch} we obtain the Fokker--Planck
equation (\ref{dFok_Plank}) with coefficients ${D^{(1)}}$ and ${D^{(2)}}$:
\begin{eqnarray}
{D^{(1)}} &=& \frac{1}{\gamma} \left\{
\sigma\left[T_e-1-\frac{1}{g}\left(\frac{\theta^{-1}+\sigma/\alpha}
{1+\sigma/\alpha}\right)\right]\right. \nonumber \\ &+&
\sigma^3\left[1-\frac{1}{g}\left(\frac{\theta^{-1}+\sigma/\alpha}
{1+\sigma/\alpha}\right) \right] - I\sigma \tau_\lambda \nonumber \\ &-& \left.
\frac{2I\sigma}{\gamma^2} \left[ \gamma - \frac{\sigma}{g} \left(
\frac{\tau_\varepsilon(1 - \theta^{-1})} {(1+\sigma/\alpha)^3\alpha} +
\sigma\tau_p\right)\right]\right\}, \label{D_1n} \\ {D^{(2)}} &=&
\frac{I\sigma^2} {\gamma} \left[\gamma^{-1} + 2\tau_\lambda \right].
\label{D_2n}
\end{eqnarray}

Using Eqs.~(\ref{extr}), (\ref{D_1n}), and (\ref{D_2n}),
the equality is found with the same meaning as (\ref{Teq35}), which gives
the boundary of existence of distribution (\ref{dpdf}) maximum
at zero value of stress corresponding to the solidlike state of lubricant
\begin{equation} \label{Teq35n}
T_{e0} = \frac{\theta^{-1}+g_\Theta}{g_\Theta}
+\left(\tau_\lambda +\frac{2g_\Theta}{\theta^{-1}\tau_\varepsilon+
\tau_p}\right)I. \end{equation}

The $P(\sigma)$ dependencies are shown in Fig.~4 for the different modes of
friction. Curves 1 -- 5 correspond to the domains of dry (DF), stick-slip (SS),
stick-slip and sliding (SS+SF), metastable and stable sliding (MSF+SF), and
sliding friction (SF), respectively.
The phase diagram and portraits are presented in
Figs.~5 and 6 corresponding to the curves of Fig.~4.

The most complex form of $P(\sigma)$ function is inherent
in SS+SF region (curve 3 in Fig.~4). Here
the solidlike, metastable and stable liquidlike lubricant states coexist
meeting the maximums of $P(\sigma)$. It means the realization possibility of
intermittent (stick-slip) friction,
at which the periodic transitions occur between
the dynamic modes corresponding to these states. In the MSF+SF domain of
stick-slip motion the metastable and stable sliding can periodically change
each other  (curve 4 in Fig.~4).
It is characteristic that transition from SS+SF to MSF+SF region is
accompanied by disappearance of dry friction in the system. The domain of dry
friction (DF) broadens, and the region of sliding friction (SF) decreases with
growth of correlation time $\tau_\lambda$ of noise.

The phase portrait of dry friction region (DF) is similar to that is
inherent in continuous transformation (Fig.~3a).
It implies that DF domain is equivalent at the taking into account of
the modulus defect and without it.

The phase portrait describing the region of stick-slip friction (SS)
is similar to the characteristic one at continuous
transformation (Fig.~3b). The basic their difference is that here
the trajectories around focus are considerably more elongated along both
coordinates axes. It means the greater stability of sliding friction.

The most complex region (SS+SF) is represented by the phase portrait
shown in Fig.~6a. The five singular points are realized here: $D$,
the saddles $N$, $N'$, the stable focuses $F$, $F'$. As well as above,
saddles correspond to the minimums of $P(\sigma)$ dependence.
Point $D$ meets the solidlike state of lubricant. Stable focus $F$
determines the first non-zero maximum of probability.
It is apparent that the oscillations are weakly pronounced
around this point. In this mode lubricant represents the very viscous
liquid, because in such type of fluid at presence of noise the
oscillations are damped strongly. Actually, the point $F$
corresponds to the small values of stress, and with its decreasing the
lubricant becomes more viscous, and it is transformed
into the solidlike state at $\sigma=0$.
Thus, using phase portraits it is possible to give explanation
to that the liquidlike state of lubricant,
but not the solidlike one, corresponds to the large values of shear stress.
Focus $F'$ meets the second non-zero maximum of
$P(\sigma)$ function, and there are oscillations with large amplitude
around it. This implies the fluidlike state of lubricant
and, accordingly, sliding. The last
point is in large distance along abscissa axis
from all others ones. This mode of friction is most probable only,
since points $D$ and $F$, corresponding to the dry and metastable
sliding friction, have large stability and probability of realization also.
From here the conclusion follows that lubricant can undergo
periodic transitions (stick-slip) between the modes corresponding
to the points $D$, $F$, and $F'$.
Since these modes are stable and separated by the pronounced
minimums of distribution function $P(\sigma)$ (by saddles), the transitions
between them is necessary to expect after large intervals of time.

The phase portrait of MSF+SF domain is represented in Fig.~6b. There are
three singular points --- the stable focuses $F$, $F'$, and the saddle $N$.
The latter is similar to the described above saddles and it meets
the minimum of probability dependence on the stress.
Point $F$ corresponds to the first maximum of distribution, which describes
metastable sliding mode (MSF), and $F'$ --- the second maximum,
which defines stable sliding (SF).
There are only insignificant oscillations around the
focus $F$, however, lubricant in this mode is less viscous
liquid than in the vicinity of the point $F'$ in Fig.~6a. At the origin of
coordinates the singular point is absent, and the dry friction is not realized.
Focus $F'$ is similar to described in Fig.~6a, however, it's
"attraction" \ domain is more stretched along both axes that means the
larger fluidity of lubricant and stability of this mode. Therefore
in comparison with the previous case, here the arising of
sliding friction is more probable.

SF region is represented by the phase portrait, which is similar to
the described for the continuous transformation (Fig.~3c).
Here the one stable focus $F$ is realized, characterized by oscillations
in it's vicinity, representing stable sliding friction.
The basic difference is that in this case
oscillations take place with large amplitude that implies strong fluidity of
lubricant and pronounced stability of such mode. However, as well as in
all above considered situations, in the course of time in lubricant
the stationary shear stress is set corresponding to the
maximum of the initial distribution $P(\sigma)$.

In the basic equations (\ref{eq2}) -- (\ref{eq4}) the shear stress
$\sigma$ stands in the first power. However, in general case it's
exponent $a$ may be fractional ($0<a<1$), but not integer:
\begin{eqnarray}
\tau _\sigma \dot {\sigma } &=& - \sigma^a + g\varepsilon , \label{eq2_a} \\
\tau _\varepsilon \dot {\varepsilon } &=& - \varepsilon + (T - 1)\sigma^a ,
\label{eq3_a} \\
\tau _T \dot {T} &=& (T_e - T) - \sigma^a \varepsilon + \sigma ^{2a}
+\lambda(t). \label{eq4_a}
\end{eqnarray}
Taking into account the additive noises of the shear stress and strain,
and the temperature of lubricant film it has been shown \cite{physa_soc}
that such system describes the self-similar mode for which the
characteristic scale of shear stress is absent \cite{Amit}.
This regime is determined by the homogeneous distribution function
\begin{equation}
P\left( y \right) = y^{- 2a} {\mathcal P}
\left( \sigma \right), \quad y=\sigma\sigma_s . \label{14}
\end{equation}
In particular, the value $2a=1.5$ corresponds to the self-organized
criticality mode, at which, unlike the phase transition,
the process of self-organization does not require the external
influence ($T_e = 0$) and occurs spontaneously \cite{physa_soc,0}.

The self-similar behaviour of lubricant film is studied taking into account
fluctuations of its temperature defined by Ornstein-Uhlenbeck process.
It is shown that the fluctuations of lubricant temperature result
in disappearance of sliding friction region at presence of dry
and stick-slip friction domains in both cases of second-order and
first-order transitions. In the second case the stick-slip motion arises
characterized by three stationary values of shear stresses at which dry,
metastable and stable sliding friction are realized. The increase
of correlation time of lubricant temperature fluctuations leads to
increasing of frictional surfaces temperature needed for realization
of stick-slip friction.

The study of Eqs.~(\ref{eq2_a}) -- (\ref{eq4_a})
shows that the phase portraits for similar domains reproduce
above considered qualitatively. However, there are substantial differences.
The fractional Lorenz system at $a \ne 1$ and $I \ne 0$
results in presence of the singular point $D$
in phase portraits, which corresponds to the solidlike state of lubricant.
Besides, the variation of $a$ leads to the complication of $P(\sigma)$
dependence, and as a result, to more complex form of phase portraits.
Within the determined friction mode at decreasing of $a$
the increase of abscissas of the stable focuses is observed.
Consequently, the weakening of
fractional feedbacks in the Lorenz-type models results
in the increase of lubricant fluidity and the reducing of friction.
However, in the systems described by fractional exponent $a$ the dry
friction is realized always. Thus, it is impossible to assert that
such systems more preferable to friction decrease than linear systems.

\section{Conclusion}\label{sec:level4}

The above consideration shows that increase of temperature of frictional
surfaces $T_e$, at presence of colored noise of lubricant temperature, can be
accompanied by self-organization of elastic and thermal fields leading to the
mode of sliding friction. Indeed, the correlation degree of lubricant
temperature change plays the substantial role. If the correlation time
$\tau_\lambda$ increases, the growth of friction surfaces temperature is
necessary for transition from dry to sliding friction mode at the fixed
intensity $I$ of temperature fluctuations. In the case of continuous
transformation at small intensity $I$ this transition takes place avoiding the
region of intermittent friction, i.e., has the form of second-order transition
--- the melting of amorphous lubricant. In the reverse case of large $I$ the
first-order transition is realized corresponding to the melting of crystalline
lubricant.

At setting of the sliding friction mode in the system the damping oscillations
arise and the shear stress relaxes to the stationary value
fixed by the probability
distribution $P(\sigma)$. The amplitude of these oscillations increases with
growth of stationary values of shear stress. It means that large shear stress
$\sigma$ corresponds to the liquidlike structure of lubricant. The solidlike
state of lubricant is described by the singular point $D$ at the origin of
coordinates that has complex character of stability and corresponds to the
divergence of probability $P(\sigma)$. The oscillations near this point are
absent.

For description of first-order transition the shear modulus
defect is taken into account. It is shown that the change of
temperature fluctuations intensity $I$ and frictional surfaces
temperature $T_e$ can transform the system from the dry friction mode
to the sliding. The latter arises at two values of shear stress.
Accordingly, the three singular points, that define the
stationary values of stress, appear in phase portraits --- non-standard
point $D$ at zero stress and two stable focuses at non-zero ones.
The intermittent (stick-slip) mode of friction can be realized as a result
of transitions between solidlike, metastable and stable liquidlike lubricant
states, which are described by zero and non-zero singular points.

Taking into consideration the nonlinear relaxation of the shear stress
and fractional feedbacks in the Lorenz system it has been shown
that the region of sliding friction is realized in the phase
diagram only in absence of temperature fluctuations.
In this case the singular point $D$ is always present in phase portraits,
which corresponds to the solidlike state of lubricant and dry
friction. Besides, the change of fractional exponent can complicate
the phase diagram, increasing the number of friction domains,
in accordance with experimental data \cite{Yosh}.

\section*{ACKNOWLEDGMENTS}

I thank I.~A.~Lyashenko and Dr.~O.V.~Boyko for attentive reading and 
correction of the manuscript. The work was partly supported by a grant of 
the cabinet of Ukraine.

\newpage

\begin{center} {\bf Figure captions} \\ to the paper by
A.~V.~Khomenko \\ "Influence of temperature correlations on phase
dynamics and kinetics of ultrathin lubricant film"
\end{center}

\begin{description}

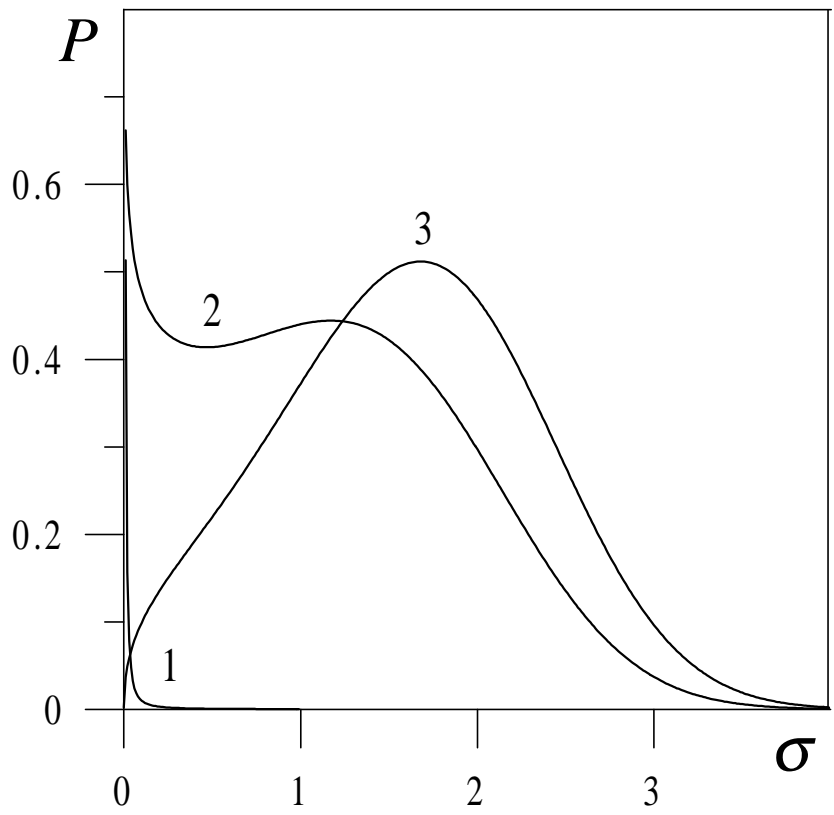
\item[FIG.~1.] The distribution function of the shear stress for
the second-order transition at
$g = 0.2,~\tau_\sigma = \tau_\varepsilon = 0.1,~\tau_\lambda = 0.2,$
and $I = 5$. The curves 1, 2, 3 correspond to the temperatures
$T_e = 5, 16, 20$, respectively.

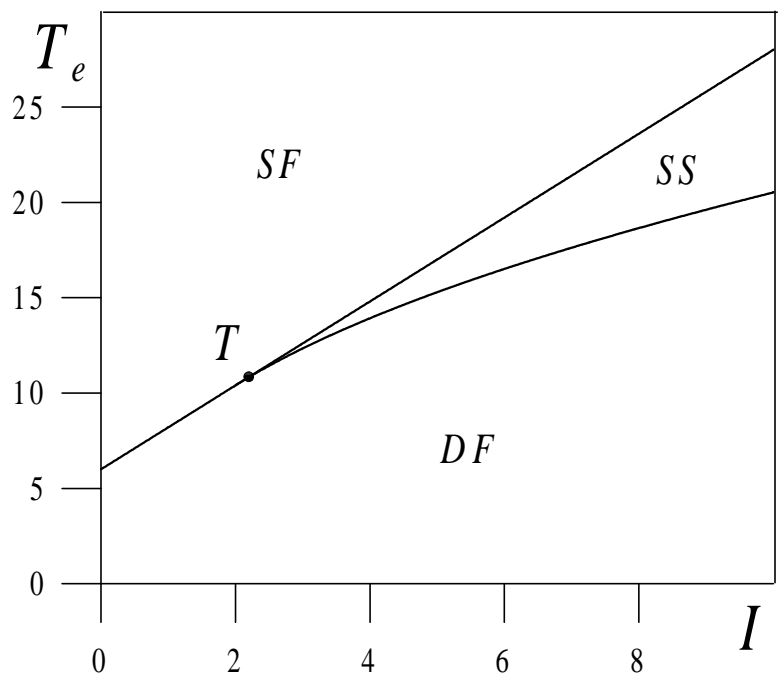
\item[FIG.~2.] The phase diagram corresponding to the parameters of Fig.~1
with the domains of dry (DF), sliding (SF), and stick-slip (SS) friction
($T$ is the tricritical point).

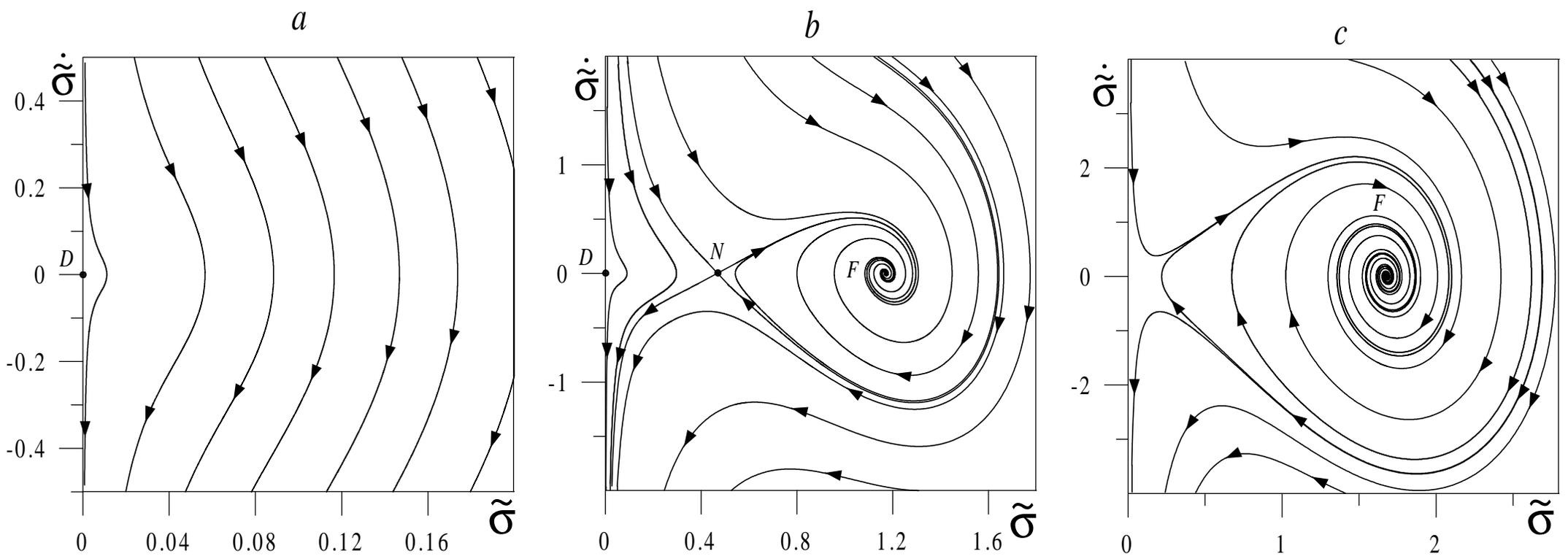
\item[FIG.~3.] The phase portraits corresponding to the parameters of
Fig.~1: (a) DF mode corresponds to the curve 1 in Fig.~1;
(b)  SS  --- curve 2 in  Fig.~1; (c) SF  --- curve 3  in  Fig.~1.

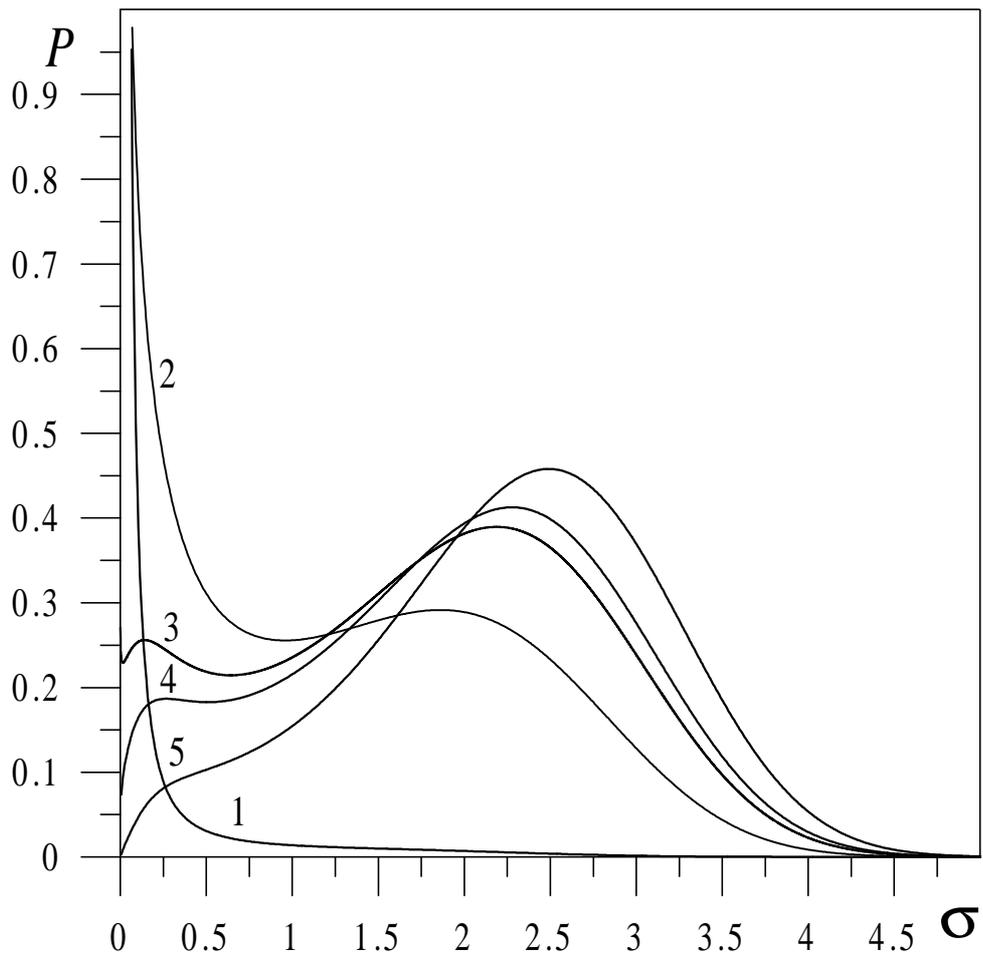
\item[FIG.~4.] The distribution function of shear stress for the
first-order transition at $\tau_p = \tau_\varepsilon = \tau_\lambda = 0.1,
~\theta^{-1} = 7, ~\alpha = 0.3, ~g_\Theta = 0.4, I = 4.5$.
The curves 1 -- 5 correspond to the temperatures
$T_e {=} 16, 21, 23.25, 24, 26$, respectively.

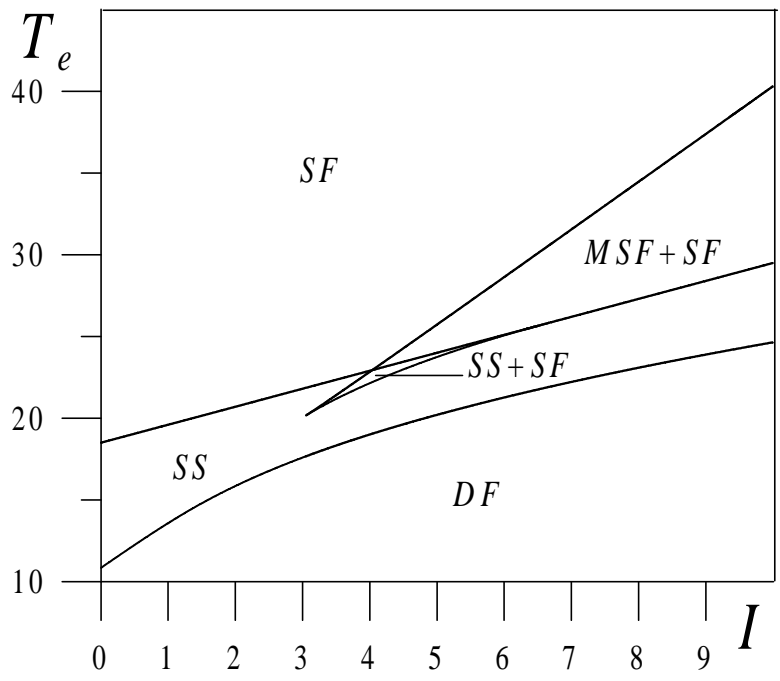
\item[FIG.~5.] The phase diagram corresponding to the parameters of Fig.~4
with the domains of dry (DF), sliding (SF), and stick-slip (SS, MSF+SF,
SS+SF) friction modes.

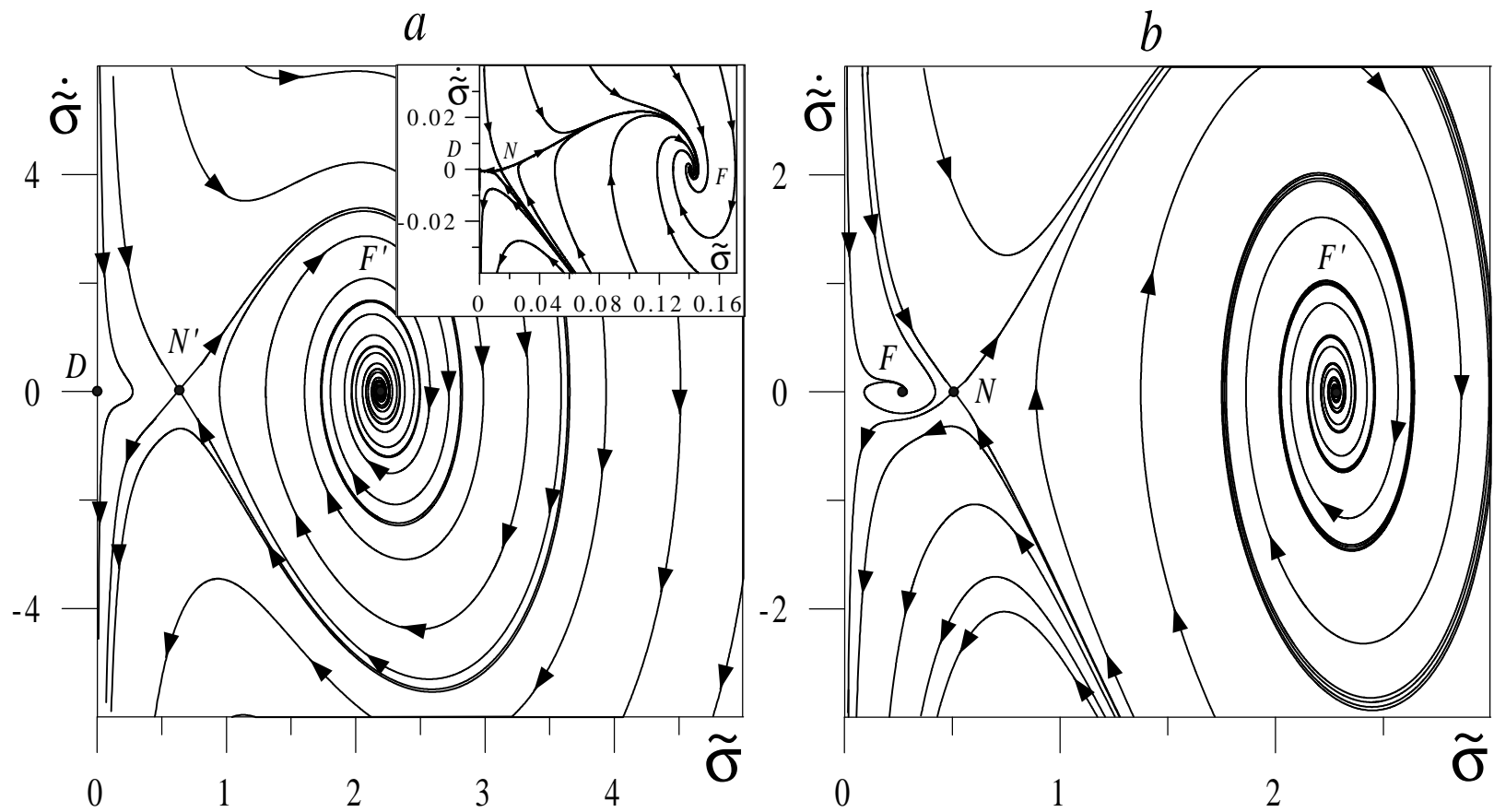
\item[FIG.~6.] The phase portraits corresponding to the parameters of Fig.~4:
(a) SS+SF mode corresponds to the curve 3 in Fig.~4;
(b) MSF+SF  --- curve 4  in Fig.~4.

\end{description}

\end{document}